\documentclass[prd,aps,preprint,showpacs,preprintnumbers,amsmath,amssymb,tighten,nofootinbib,12pt]{revtex4}

\usepackage{graphicx}
\usepackage{dcolumn}
\usepackage{bm}

\begin{document}


\title{Radiative Corrections to the CKM  Unitarity Triangles
\vspace{0.3cm}}

\author{\bf Shu Luo}
\email{luoshu@ihep.ac.cn}

\author{\bf Zhi-zhong Xing}
\email{xingzz@ihep.ac.cn}

\affiliation{Institute of High Energy Physics
and Theoretical Physics Center for Science Facilities, \\ Chinese
Academy of Sciences, P.O. Box 918, Beijing 100049, China\vspace{0.7cm}}

\begin{abstract}
The $3\times 3$ CKM matrix defines six unitarity triangles in the
complex plane, which will be carefully explored in the LHCb
experiment and at the Super-$B$ factory. We calculate the running
effects of nine different inner angles and eighteen different sides
of the six triangles from the electroweak scale to a
superhigh-energy scale by using the one-loop renormalization-group
equations, and demonstrate that all the nine angles are stable
against radiative corrections. In particular, we find that the
CP-violating angle $\alpha$ is most insensitive to the changes of
energy scales.
\end{abstract}


\maketitle

The 10-year running of the BARBAR and Belle $B$-meson factories has
greatly improved our knowledge on the Cabibbo-Kobayashi-Maskawa (CKM)
matrix $V$ \cite{PDG08}, which describes the flavor-changing effects in weak
charged-current interactions of $(u, c, t)$ and $(d, s, b)$ quarks.
In particular, the observed CP-violating asymmetries in a
number of $B$ decays have unambiguously and consistently verified the
Kobayashi-Maskawa (KM) mechanism of CP violation in the standard model (SM).
More extensive and precise studies of quark flavor mixing and
CP violation will be done definitely in the LHCb experiment \cite{LHCb}
and hopefully at the super-$B$ factory \cite{SB}, from which one may
test the KM picture to an unprecedented degree of accuracy.

Theorists expect that the observed flavor puzzles at low energies,
such as the strong hierarchies of quark masses and flavor mixing
angles, should be resolved in a predictive flavor theory at a
superhigh-energy scale $\Lambda$. Such a theory can be confronted
with the experimental data via the renormalization-group equations
(RGEs) which run its predictions from $\Lambda$ down to the
electroweak scale. One may phenomenologically do the opposite ---
running the parameters of quark flavor mixing and CP violation from
the electroweak scale up to $\Lambda$ and examining their
sensitivities to radiative corrections. A comparison between
theoretical predictions and experimental measurements will be
available once the wanted flavor theory is in hand. So far a lot of
works have been done to investigate the RGE running effects of the
CKM matrix elements with or without the help of a specific
parametrization \cite{FX00}.

The present paper aims to analyze how the unitarity triangles (UTs)
of the CKM matrix $V$, both their sides and their angles, evolve
with the energy scales. This analysis makes sense because the
precision measurements to be done in the LHCb experiments and at the
super-$B$ factory will probe all the six UTs or at least several of
them, in order to test the KM mechanism of CP violation and explore
possible new physics beyond it. We derive the one-loop RGEs for the
$3\times 3$ CKM angle matrix $\Phi$ proposed recently by Harrison
{\it et al.} \cite{Harrison}, and demonstrate that its nine angles
are all stable against radiative corrections. In particular, we find
that the CP-violating angle $\Phi^{}_{cs} =\alpha$ is most
insensitive to the RGE running effect. We also present the one-loop
RGEs for the eighteen sides of six UTs and for the Jarlskog
invariant of CP violation. The running behaviors of these quantities
are numerically illustrated by assuming $\Lambda \sim 10^{14}$ GeV,
which is very close to the scale of grand unified theories or to the
scale of conventional seesaw mechanisms in the framework of either
the SM or the minimal supersymmetric standard model (MSSM). Our
results are expected to be helpful for building quantitatively
viable flavor models at superhigh-energy scales.

\vspace{0.3cm}

Because the $3\times 3$ CKM matrix $V$ is unitary, its nine elements
satisfy the following normalization and orthogonality conditions:
\begin{equation}
\sum_\alpha V^{}_{\alpha i} V^*_{\alpha j}
\; =\; \delta^{}_{ij} \; , ~~~~
\sum_i V^{}_{\alpha i} V^*_{\beta i} \; =\; \delta^{}_{\alpha\beta} \; ,
\end{equation}
where the Greek and Latin subscripts run over $(u, c, t)$ and $(d,
s, b)$, respectively. The six orthogonality relations geometrically
define six UTs in the complex plane, as illustrated in Fig. 1. The
six UTs have eighteen different sides and nine different inner
angles \cite{FX00}, but their areas are all identical to $\cal J$/2
with $\cal J$ being the Jarlskog invariant of CP violation \cite{J}
defined through
\begin{equation}
{\rm Im} \left( V^{}_{\alpha i} V^{}_{\beta j} V^*_{\alpha j} V^*_{\beta i}
\right) \; =\; {\cal J} \sum_\gamma \epsilon^{}_{\alpha\beta\gamma}
\sum_k \epsilon^{}_{ijk} \; .
\end{equation}
Nine inner angles of the CKM UTs can be defined as
\begin{equation}
\Phi^{}_{\alpha i} \; \equiv \; \arg \left(- \frac{V^{}_{\beta j}
V^*_{\gamma j}}{V^{}_{\beta k} V^*_{\gamma k}} \right) \; ,
\end{equation}
where $\alpha$, $\beta$ and $\gamma$ run co-cyclically over $u$, $c$ and $t$,
while $i$, $j$ and $k$ run co-cyclically over $d$, $s$ and $b$. Then one
may write out the CKM angle matrix \cite{Harrison}
\begin{equation}
\Phi \; = \;\left ( \begin{matrix} \Phi^{}_{ud} & ~\Phi^{}_{us}~ & \Phi^{}_{ub}
\cr \Phi^{}_{cd} & \Phi^{}_{cs} & \Phi^{}_{cb} \cr \Phi^{}_{td} &
\Phi^{}_{ts} & \Phi^{}_{tb} \end{matrix} \right ) \; .
\end{equation}
Note that the angle $\Phi^{}_{\alpha i}$ is just the inner angle
shared by the UTs $\triangle^{}_{\alpha}$ (for $\alpha = u$, $c$ or
$t$) and $\triangle^{}_{i}$ (for $i=d$, $s$ or $b$), as one can
easily see in Fig. 1. Hence each row or column of the CKM angle
matrix $\Phi$ corresponds to one UT, and its three matrix elements
satisfy the normalization conditions
\begin{equation}
\sum_\alpha \Phi^{}_{\alpha i} \; =\; \sum_i \Phi^{}_{\alpha i}
\; =\; \pi \; .
\end{equation}
This result implies that one can have two off-diagonal asymmetries of
$\Phi$ about its $\Phi^{}_{ud}$-$\Phi^{}_{cs}$-$\Phi^{}_{tb}$
and $\Phi^{}_{ub}$-$\Phi^{}_{cs}$-$\Phi^{}_{td}$ axes, respectively
\footnote{The off-diagonal asymmetries of the CKM matrix $V$, given as
$\Delta^{}_{\rm L} \equiv |V^{}_{us}|^2 - |V^{}_{cd}|^2 =
|V^{}_{cb}|^2 - |V^{}_{ts}|^2 = |V^{}_{td}|^2 - |V^{}_{ub}|^2$ and
$\Delta^{}_{\rm R} \equiv |V^{}_{us}|^2 - |V^{}_{cb}|^2 =
|V^{}_{cd}|^2 - |V^{}_{ts}|^2 = |V^{}_{tb}|^2 - |V^{}_{ud}|^2$,
have been discussed in Ref. \cite{Xing95}. We are able to write down the relations between these two types of off-diagonal asymmetries: $\sin {\cal A}^{}_{\rm L} = \displaystyle \frac{- \Delta^{}_{\rm L} {\cal J}}{|V^{}_{us}| |V^{}_{ub}| |V^{}_{cd}| |V^{}_{cb}| |V^{}_{td}| |V^{}_{ts}|}$ and $\sin {\cal A}^{}_{\rm R} = \displaystyle \frac{- \Delta^{}_{\rm R} {\cal J}}{|V^{}_{ud}| |V^{}_{us}| |V^{}_{cd}| |V^{}_{cb}| |V^{}_{ts}| |V^{}_{tb}|}$ .}:
\begin{eqnarray}
{\cal A}^{}_{\rm L} \; \equiv \; \Phi^{}_{us} - \Phi^{}_{cd} \; =\;
\Phi^{}_{cb} - \Phi^{}_{ts} = \Phi^{}_{td} - \Phi^{}_{ub} \; ;
\nonumber \\
{\cal A}^{}_{\rm R} \; \equiv \; \Phi^{}_{us} - \Phi^{}_{cb} \; =\;
\Phi^{}_{cd} - \Phi^{}_{ts} = \Phi^{}_{tb} - \Phi^{}_{ud} \; .
\end{eqnarray}
If ${\cal A}^{}_{\rm L} =0$ or ${\cal A}^{}_{\rm R} =0$ held,
we would be left with three pairs of congruent UTs
\footnote{Note that the unitarity of the CKM matrix $V$ allows one
to determine the moduli $|V^{}_{\alpha i}|$ in terms of the angles
$\Phi^{}_{\alpha i}$ (for $\alpha = u, c, t$ and $i = d, s, b$)
\cite{Wu}. Hence every pair of the UTs in Eq. (7) would not only
be similar to each other but also be congruent with each other, if
${\cal A}^{}_{\rm L} =0$ or ${\cal A}^{}_{\rm R} =0$ were given.}:
\begin{eqnarray}
{\cal A}^{}_{\rm L} \; =\; 0 \;\;\; & \Longrightarrow \;\;\; &
\triangle^{}_u \cong \triangle^{}_d \; , ~~
\triangle^{}_c \cong \triangle^{}_s \; , ~~
\triangle^{}_t \cong \triangle^{}_b \; ; \nonumber \\
{\cal A}^{}_{\rm R} \; =\; 0 \;\;\; & \Longrightarrow \;\;\; &
\triangle^{}_u \cong \triangle^{}_b \; , ~~
\triangle^{}_c \cong \triangle^{}_s \; , ~~
\triangle^{}_t \cong \triangle^{}_d \; .
\end{eqnarray}
Interestingly, the same expectation would be true if one of the
off-diagonal asymmetries of the CKM matrix $V$ itself (i.e.,
$\Delta^{}_{\rm L}$ or $\Delta^{}_{\rm R}$) were exactly vanishing
\cite{Xing95}.

In Ref. \cite{Harrison} the notations of $\Phi^{}_{\alpha i}$ have
been linked to the conventional ones used to denote the inner angles
of the most popular UT $\triangle^{}_s$ and some other CP-violating
phases: $\Phi^{}_{ud} = \beta^{}_s = \chi$, $\Phi^{}_{us} = \beta =
\phi^{}_1$, $\Phi^{}_{cd} = \gamma^\prime = \gamma -\delta \gamma$,
$\Phi^{}_{cs} = \alpha = \phi^{}_2$, $\Phi^{}_{cb} = \beta +
\delta\gamma$, $\Phi^{}_{ts} = \gamma = \phi^{}_3$ and $\Phi^{}_{tb}
= \beta^{}_K = \chi^\prime$. The present experimental data on CP
violation yield
\begin{equation}
\Phi \; =\; \left( \begin{matrix} 1.04^\circ \pm 0.05^\circ &
21.58^\circ \pm 0.86^\circ & 157.38^\circ \mp 0.89^\circ \cr
66.82^\circ \mp 4.20^\circ & 90.60^\circ \pm 4.00^\circ &
22.58^\circ \pm 0.89^\circ \cr 112.14^\circ \pm 4.21^\circ &
67.82^\circ \mp 4.22^\circ & 0.035^\circ \pm 0.003^\circ \cr
\end{matrix} \right) \; ,
\end{equation}
where the normalization conditions in Eq. (5) have been used
\cite{Harrison}. We can therefore obtain ${\cal A}^{}_{\rm L} =
\beta - \gamma + \delta\gamma \approx -45^\circ$ and ${\cal
A}^{}_{\rm R} = -\delta\gamma \approx -1^\circ$, in contrast with
$\Delta^{}_{\rm L} \approx 6.4 \times 10^{-5}$ and $\Delta^{}_{\rm
R} \approx 5.1 \times 10^{-2}$. In other words, the CKM matrix $V$
is almost symmetric about its $V^{}_{ud}$-$V^{}_{cs}$-$V^{}_{tb}$
axis, while the CKM angle matrix $\Phi$ is approximately symmetric
about its $\Phi^{}_{ub}$-$\Phi^{}_{cs}$-$\Phi^{}_{td}$ axis. The
nine angles $\Phi^{}_{\alpha i}$ signify different CP-violating
effects in a variety of $B$-, $D$- and $K$-meson decay modes
\cite{Wu,Bigi}. So it is very  desirable to determine $\Phi$ as
precisely as possible in the upcoming LHCb experiment and at the
future super-$B$ factory, in order to precisely test the CKM
unitarity. In this sense we argue that the CKM angle matrix $\Phi$
is a useful phenomenological language to describe CP violation.

We proceed to derive the one-loop RGEs of the CKM angle matrix $\Phi$.
The one-loop RGEs of the gauge couplings and charged-lepton and quark
Yukawa couplings have already been calculated by several
authors \cite{RGE}
\footnote{We neglect the tiny RGE effects associated with three neutrinos
by assuming that possible heavy degrees of freedom responsible for
their mass generation (e.g., heavy particles in the seesaw
mechanisms \cite{SS}) are decoupled at $\Lambda$ and thus do not
affect the RGEs of charged-lepton and quark Yukawa couplings.}.
Here we make use of their results for the RGEs of the CKM matrix
elements $|V^{}_{\alpha i}|^2$ by taking account of $y^2_u \ll
y^2_c \ll y^2_t$ and $y^2_d \ll y^2_s \ll y^2_b$, where
$y^{}_\alpha$ and $y^{}_i$ stand respectively for the eigenvalues
of the Yukawa coupling matrices of up- and down-type quarks. In
this excellent approximation we have \cite{Xing09}
\begin{eqnarray}
& & 16\pi^2 \frac{\rm d}{{\rm d}t} \left[ \begin{matrix} |V^{}_{ud}|^2 &
|V^{}_{us}|^2 & |V^{}_{ub}|^2 \cr |V^{}_{cd}|^2 & |V^{}_{cs}|^2 &
|V^{}_{cb}|^2 \cr |V^{}_{td}|^2 & |V^{}_{ts}|^2 & |V^{}_{tb}|^2
\cr \end{matrix} \right ] \nonumber \\
& = & 2C y^2_b \left[
\begin{matrix} |V^{}_{ud}|^2 |V^{}_{ub}|^2 & |V^{}_{us}|^2 |V^{}_{ub}|^2
& -|V^{}_{ub}|^2 \left( 1 - |V^{}_{ub}|^2 \right) \cr
|V^{}_{td}|^2 |V^{}_{tb}|^2 - |V^{}_{ud}|^2 |V^{}_{ub}|^2 &
|V^{}_{ts}|^2 |V^{}_{tb}|^2 - |V^{}_{us}|^2 |V^{}_{ub}|^2 &
-|V^{}_{cb}|^2 \left( |V^{}_{tb}|^2 - |V^{}_{ub}|^2 \right) \cr
-|V^{}_{td}|^2 |V^{}_{tb}|^2 & -|V^{}_{ts}|^2 |V^{}_{tb}|^2 &
|V^{}_{tb}|^2 \left( 1 -
|V^{}_{tb}|^2 \right) \cr \end{matrix} \right] \nonumber \\
& + & 2C y^2_t \left[ \begin{matrix} |V^{}_{ud}|^2 |V^{}_{td}|^2 &
|V^{}_{ub}|^2 |V^{}_{tb}|^2 - |V^{}_{ud}|^2 |V^{}_{td}|^2 &
-|V^{}_{ub}|^2 |V^{}_{tb}|^2 \cr |V^{}_{cd}|^2 |V^{}_{td}|^2 &
|V^{}_{cb}|^2 |V^{}_{tb}|^2 - |V^{}_{cd}|^2 |V^{}_{td}|^2 &
-|V^{}_{cb}|^2 |V^{}_{tb}|^2 \cr -|V^{}_{td}|^2 \left( 1 -
|V^{}_{td}|^2 \right) & -|V^{}_{ts}|^2 \left( |V^{}_{tb}|^2 -
|V^{}_{td}|^2 \right) & |V^{}_{tb}|^2 \left( 1 - |V^{}_{tb}|^2
\right) \cr \end{matrix} \right] \; ,
\end{eqnarray}
where $t\equiv \ln (\mu /M^{}_Z)$, $C=-1.5$ in the SM and $C=+1$
in the MSSM. We first derive the RGE of $\Phi^{}_{cs}$. This angle
is related to the sides of the UT $\triangle^{}_{s}$ through the
cosine rule $2 |V^{}_{ub}| |V^{}_{ud}| |V^{}_{tb}| |V^{}_{td}|
\cos\Phi^{}_{cs} = |V^{}_{ub}|^2 |V^{}_{ud}|^2 +
|V^{}_{tb}|^2 |V^{}_{td}|^2 - |V^{}_{cb}|^2 |V^{}_{cd}|^2$. Therefore,
\begin{eqnarray}
\frac{\rm d}{{\rm d}t} \cos\Phi^{}_{cs} & = &
\frac{\rm d}{{\rm d}t} \left ( \frac{|V^{}_{ub}|^2 |V^{}_{ud}|^2 +
|V^{}_{tb}|^2 |V^{}_{td}|^2 -
|V^{}_{cb}|^2 |V^{}_{cd}|^2}{2 |V^{}_{ub}| |V^{}_{ud}| |V^{}_{tb}| |V^{}_{td}|}
\right ) \nonumber \\
& = & \frac{1}{4} \left( \frac{|V^{}_{ub}| |V^{}_{ud}|}{|V^{}_{tb}| |V^{}_{td}|} -
\frac{|V^{}_{tb}| |V^{}_{td}|}{|V^{}_{ub}| |V^{}_{ud}|} \right )
\left ( \frac{1}{|V^{}_{ub}|^2} \frac{\rm d}{{\rm d}t} |V^{}_{ub}|^2 +
\frac{1}{|V^{}_{ud}|^2} \frac{\rm d}{{\rm d}t} |V^{}_{ud}|^2 -
\frac{1}{|V^{}_{tb}|^2} \frac{\rm d}{{\rm d}t} |V^{}_{tb}|^2
\right . \nonumber \\
&& \left . - \frac{1}{|V^{}_{td}|^2} \frac{\rm d}{{\rm d}t} |V^{}_{td}|^2 \right )
+ \frac{|V^{}_{cb}|^2 |V^{}_{cd}|^2}{4|V^{}_{ub}| |V^{}_{ud}| |V^{}_{tb}| |V^{}_{td}|}
\left ( \frac{1}{|V^{}_{ub}|^2} \frac{\rm d}{{\rm d}t} |V^{}_{ub}|^2 +
\frac{1}{|V^{}_{ud}|^2} \frac{\rm d}{{\rm d}t} |V^{}_{ud}|^2 \right . \nonumber \\
&& \left . + \frac{1}{|V^{}_{tb}|^2} \frac{\rm d}{{\rm d}t}
|V^{}_{tb}|^2 + \frac{1}{|V^{}_{td}|^2} \frac{\rm d}{{\rm d}t}
|V^{}_{td}|^2 - \frac{2}{|V^{}_{cb}|^2} \frac{\rm d}{{\rm d}t}
|V^{}_{cb}|^2 - \frac{2}{|V^{}_{cd}|^2} \frac{\rm d}{{\rm d}t}
|V^{}_{cd}|^2 \right ) \; .
\end{eqnarray}
Substituting the relevant expressions of ${\rm d}|V^{}_{\alpha i}|^2/{\rm d}t$
given in Eq. (9) into the right-hand side of Eq. (10), we immediately arrive at
\begin{equation}
16\pi^2 \frac{\rm d}{{\rm d}t} \Phi^{}_{cs} \; = \;
-16\pi^2 \frac{1}{\sin\Phi^{}_{cs}}
\frac{\rm d}{{\rm d}t} \cos\Phi^{}_{cs} \; = \;0 \; .
\end{equation}
This result is consistent with ${\rm d}\alpha/{\rm d}t =0$ obtained
in Ref. \cite{Xing09} from a slightly different way. With the help
of Eqs. (9) and (11), we may easily calculate the RGE of the
Jarlskog invariant ${\cal J}$ from the relation ${\cal J} =
|V^{}_{ud}| |V^{}_{ub}| |V^{}_{td}| |V^{}_{tb}| \sin\Phi^{}_{cs}$:
\begin{eqnarray}
16\pi^2 \frac{\rm d}{{\rm d}t} {\cal J} & = & 16\pi^2 \sin\Phi^{}_{cs}
\frac{\rm d}{{\rm d}t} \left ( |V^{}_{ud}|  |V^{}_{ub}|  |V^{}_{td}|
|V^{}_{tb}| \right ) \nonumber\\
& = & 8\pi^2 {\cal J} \left ( \frac{1}{|V^{}_{ub}|^2} \frac{\rm d}{{\rm d}t}
|V^{}_{ub}|^2 + \frac{1}{|V^{}_{ud}|^2} \frac{\rm d}{{\rm d}t} |V^{}_{ud}|^2
+ \frac{1}{|V^{}_{tb}|^2} \frac{\rm d}{{\rm d}t} |V^{}_{tb}|^2 +
\frac{1}{|V^{}_{td}|^2} \frac{\rm d}{{\rm d}t} |V^{}_{td}|^2 \right )
~~~~~ \nonumber\\
& = & - \; 2 C {\cal J} \left [ y^{2}_{b} \left ( |V^{}_{tb}|^2 -
|V^{}_{ub}|^2 \right ) + y^{2}_{t} \left ( |V^{}_{tb}|^2 -
|V^{}_{td}|^2 \right ) \right ] \; .
\end{eqnarray}
The one-loop RGEs for other angles of the CKM UTs can then be
derived from Eqs. (9) and (12) by using the sine rule (and the
cosine rule) repeatedly. For instance, the relationship ${\cal J} =
|V^{}_{cs}|  |V^{}_{cb}|  |V^{}_{ts}|  |V^{}_{tb}| \sin\Phi^{}_{ud}
= |V^{}_{cd}|  |V^{}_{cb}|  |V^{}_{td}|  |V^{}_{tb}|
\sin\Phi^{}_{us} = |V^{}_{us}|  |V^{}_{ub}|  |V^{}_{ts}|
|V^{}_{tb}| \sin\Phi^{}_{cd}$ allows us to separately calculate the
RGEs of $\Phi^{}_{ud}$, $\Phi^{}_{us}$ and $\Phi^{}_{cd}$. We find
\begin{eqnarray}
16\pi^2 \frac{\rm d}{{\rm d}t} \Phi^{}_{ud} & = & - \;
2 C \left ( y^{2}_{b} + y^{2}_{t} \right )
\frac{\cal J}{|V^{}_{cs}|^2} \; , \nonumber \\
16\pi^2 \frac{\rm d}{{\rm d}t} \Phi^{}_{us} & = & - \;
2 C y^{2}_{b} \frac{\cal J}{|V^{}_{cd}|^2} \; , \nonumber \\
16\pi^2 \frac{\rm d}{{\rm d}t} \Phi^{}_{cd} & = & - \;
2 C y^{2}_{t} \frac{\cal J}{|V^{}_{us}|^2} \; .
\end{eqnarray}
Taking account of Eq. (5) together with Eqs. (11) and (13), one may
simply figure out the RGEs for the other five angles of $\Phi$. Our
main results for the RGEs of $\Phi$ are summarized as
\begin{eqnarray}
16\pi^2 \frac{\rm d}{{\rm d}t} \left ( \begin{matrix} \Phi^{}_{ud} &
\Phi^{}_{us} & \Phi^{}_{ub} \cr \Phi^{}_{cd} & \Phi^{}_{cs} &
\Phi^{}_{cb} \cr \Phi^{}_{td} & \Phi^{}_{ts} & \Phi^{}_{tb}
\end{matrix} \right ) & = & 2 C {\cal J} \left [
\frac{y^{2}_{b}}{|V^{}_{cd}|^2 |V^{}_{cs}|^2} \left (
\begin{matrix} - |V^{}_{cd}|^2   & - |V^{}_{cs}|^2  & 1 -
|V^{}_{cb}|^2  \cr 0 & 0 & 0 \cr  |V^{}_{cd}|^2  & |V^{}_{cs}|^2 &
|V^{}_{cb}|^2 - 1 \end{matrix} \right )
\right. \nonumber\\
& & \left. ~~~~~ +  \frac{y^{2}_{t}}{|V^{}_{us}|^2 |V^{}_{cs}|^2}
\left (  \begin{matrix} - |V^{}_{us}|^2 & \;~~0~~\; & |V^{}_{us}|^2
\cr - |V^{}_{cs}|^2 & \;~~0~~\; & |V^{}_{cs}|^2 \cr 1 -
|V^{}_{ts}|^2 & \;~~0~~\; & |V^{}_{ts}|^2 - 1 \end{matrix} \right )
\right ] \; . ~~~~~
\end{eqnarray}
Some discussions are in order.
\begin{itemize}
\item     The angle $\Phi^{}_{cs} = \alpha$ is most insensitive to
the RGE running effect. As shown in Eq. (8), current experimental
data point to $\Phi^{}_{cs} = \alpha =\pi/2$ to an excellent degree
of accuracy, implying that the UTs $\triangle^{}_c$ and
$\triangle^{}_s$ are almost the right triangles. This possibility
was conjectured long time ago in an attempt to explore the realistic
textures of quark mass matrices \cite{FX95}. Some interest has
recently been paid to whether there is a good reason for
$\Phi^{}_{cs} = \alpha =\pi/2$ in understanding quark flavor mixing
and CP violation \cite{Harrison,Xing09,Others}.

\item     Other angles of $\Phi$ receive negligibly small radiative
corrections. $\Phi^{}_{tb}$ should be most sensitive to the RGE
running effect: $16\pi^2 {\rm d}\Phi^{}_{tb}/{\rm d}t \approx -2 C
{\cal J} (y^2_b + y^2_t)/|V^{}_{us}|^2$ holds in the approximations
of $|V^{}_{cd}| \approx |V^{}_{us}|$, $|V^{}_{cs}| \approx 1$ and
$|V^{}_{cb}|^2 -1 \approx |V^{}_{ts}|^2 -1 \approx -1$. Considering
${\cal J} \approx 3.0 \times 10^{-5}$, $|V^{}_{us}|^2 \approx 5.1
\times 10^{-2}$ and $y^2_b \lesssim y^2_t \sim {\cal O}(1)$ in the
SM or MSSM, we can roughly obtain $|\Phi^{}_{tb}(\Lambda) -
\Phi^{}_{tb}(M^{}_Z)| \lesssim {\cal O}(10^{-5}) \ln
(\Lambda/M^{}_Z)$. Hence the RGE correction to $\Phi^{}_{tb}$ is
expected to be of ${\cal O}(10^{-4})$ for $\Lambda \sim 10^{16}$ GeV.
\end{itemize}
The stability of $\Phi^{}_{\alpha i}$ (for $\alpha = u, c, t$ and $i
= d, s, b$) against radiative corrections indicates that the shape
of every UT is almost unchanged from $M^{}_Z$ to $\Lambda$ or vice
versa. So one may directly confront the experimental data on these
CP-violating angles at low energies with the predictions from a
superhigh-energy flavor model.

Although the shape of each UT is insensitive to the changes of
energy scales, its three sides may not be so. With the help of Eq.
(9), it is straightforward to obtain the following approximate RGEs
for eighteen sides of the six UTs:
\begin{eqnarray}
\triangle^{}_{u}: & & \frac{\rm d}{{\rm d}t} \ln |V^{}_{cd}
V^{*}_{td}| \; \approx \; \frac{\rm d}{{\rm d}t} \ln |V^{}_{cs}
V^{*}_{ts}| \; \approx\; \frac{\rm d}{{\rm d}t} \ln |V^{}_{cb}
V^{*}_{tb}| \; \approx \; - \; \frac{C \left ( y^{2}_{b} + y^{2}_{t}
\right )}{16\pi^2} \; , \nonumber \\
\triangle^{}_{c}: & & \frac{\rm d}{{\rm d}t} \ln |V^{}_{ud}
V^{*}_{td}| \; \approx \; \frac{\rm d}{{\rm d}t} \ln |V^{}_{us}
V^{*}_{ts}| \; \approx\; \frac{\rm d}{{\rm d}t} \ln |V^{}_{ub}
V^{*}_{tb}| \; \approx \; - \; \frac{C \left ( y^{2}_{b} + y^{2}_{t}
\right )}{16\pi^2} \; , \nonumber \\
\triangle^{}_{t}: & &  \frac{\rm d}{{\rm d}t} \ln |V^{}_{ud}
V^{*}_{cd}| \; \approx\; \frac{\rm d}{{\rm d}t} \ln |V^{}_{us}
V^{*}_{cs}| \; \approx \; 0 \; , ~~~~ \frac{\rm d}{{\rm d}t} \ln
|V^{}_{ub} V^{*}_{cb}| \; \approx \; - \; \frac{2 C \left (
y^{2}_{b} + y^{2}_{t} \right )}{16\pi^2} \; , \nonumber \\
\triangle^{}_{d}: & & \frac{\rm d}{{\rm d}t} \ln |V^{}_{us}
V^{*}_{ub}| \; \approx \; \frac{\rm d}{{\rm d}t} \ln |V^{}_{cs}
V^{*}_{cb}| \; \approx\; \frac{\rm d}{{\rm d}t} \ln |V^{}_{ts}
V^{*}_{tb}| \; \approx \; - \; \frac{C \left ( y^{2}_{b} + y^{2}_{t}
\right )}{16\pi^2} \; , \nonumber \\
\triangle^{}_{s}: & & \frac{\rm d}{{\rm d}t} \ln |V^{}_{ud}
V^{*}_{ub}| \; \approx \; \frac{\rm d}{{\rm d}t} \ln |V^{}_{cd}
V^{*}_{cb}| \; \approx\; \frac{\rm d}{{\rm d}t} \ln |V^{}_{td}
V^{*}_{tb}| \; \approx \; - \; \frac{C \left ( y^{2}_{b} + y^{2}_{t}
\right )}{16\pi^2} \; , \nonumber \\
\triangle^{}_{b}: & & \frac{\rm d}{{\rm d}t} \ln |V^{}_{ud}
V^{*}_{us}| \; \approx \; \frac{\rm d}{{\rm d}t} \ln |V^{}_{cd}
V^{*}_{cs}| \; \approx \; 0 \; , ~~~~ \frac{\rm d}{{\rm d}t} \ln
|V^{}_{td} V^{*}_{ts}| \; \approx \; - \; \frac{2 C \left (
y^{2}_{b} + y^{2}_{t} \right )}{16\pi^2} \; . ~~~~~~
\end{eqnarray}
This result is apparently consistent with the one obtained in Eq.
(14) for nine angles of the UTs. For example, three sides of
$\triangle^{}_u$ run with energy scales in the same way, and thus
its three inner angles keep unchanged from $M^{}_Z$ to $\Lambda$ or
vice versa. The UTs $\triangle^{}_c$, $\triangle^{}_d$ and
$\triangle^{}_s$ have the same RGE running behaviors as
$\triangle^{}_u$ does. As for the UT $\triangle^{}_t$ or
$\triangle^{}_b$, its two long sides are stable against radiative
corrections but its short side slightly changes with energy scales.
Because the ratio of the short side to one of the long sides is of
${\cal O}(10^{-3})$ for either $\triangle^{}_t$ or $\triangle^{}_b$,
in accordance with its smallest inner angle $\Phi^{}_{tb} \approx
0.035^\circ$ at $M^{}_Z$, the slight running of the short side has
little effect on three inner angles.

To illustrate, we use the central values of the CKM angle matrix
elements in Eq. (8) to calculate the RGE running effects of the
Jarlskog invariant ${\cal J}$ and eighteen sides of six UTs from
$M^{}_Z$ up to $\Lambda \sim 10^{14}$ GeV. Our numerical results are
shown in Fig. 2 and Table 1, where the Higgs mass $m^{}_H = 140$ GeV
in the SM and the parameter $\tan\beta = 10$ or $50$ in the MSSM
have typically been input. One can imagine that the shape of every
UT expands for $\Lambda > M^{}_Z$ in the SM, such that both its area
($={\cal J}/2$) and sides become larger and larger when the energy
scale increases. The running behaviors of six UTs are opposite in
the MSSM: both the magnitude of $\cal J$ and those of eighteen sides
become smaller and smaller when the energy scale increases. But the
nine angles of $\Phi$ are rather stable against radiative
corrections either in the SM or in the MSSM. We have numerically
confirmed that the values of $\Phi^{}_{\alpha i}$ given in Eq. (8)
keep unchanged even when the energy scale goes to $\Lambda \sim
10^{14}$ or higher
\footnote{We find that this observation is also true when using
the two-loop RGEs of charged-lepton and quark Yukawa couplings
to do the numerical calculations \cite{XZZ08}.}.

In summary, we have calculated the one-loop RGEs for the CKM angle
matrix $\Phi$ and the sides of six UTs. We have clearly demonstrated
that all the nine angles of $\Phi$ are insensitive to radiative
corrections. In particular, the angle $\Phi^{}_{cs} = \alpha$ is
most insensitive to the changes of energy scales. As for the UTs
$\triangle^{}_u$, $\triangle^{}_c$, $\triangle^{}_d$ and
$\triangle^{}_s$, we find that their sides have the same RGE running
behaviors which are more or less different from those of the two
sharp-angled UTs $\triangle^{}_t$ and $\triangle^{}_b$. Our results
convincingly indicate that the experimental data on nine
CP-violating angles $\Phi^{}_{\alpha i}$ (for $\alpha =u, c, t$ and
$i=d, s, b$), which will be well measured in the upcoming LHCb
experiment and at the future super-$B$ factory, can directly be
confronted with the predictions from a superhigh-energy flavor
theory.

\vspace{0.9cm}

We would like to thank P.F. Harrison and W.G. Scott for interesting discussions. Our research is supported in part
by the Ministry of Science and Technology of China under grant No.
2009CB825207, and in part by the National Natural Science Foundation
of China under grant No. 10425522 and No. 10875131.

\newpage

\begin{figure}
\begin{center}
\vspace{12cm}
\includegraphics[bbllx=6.5cm, bblly=6.0cm, bburx=15.0cm, bbury=14.2cm,%
width=7cm, height=7cm, angle=0, clip=0]{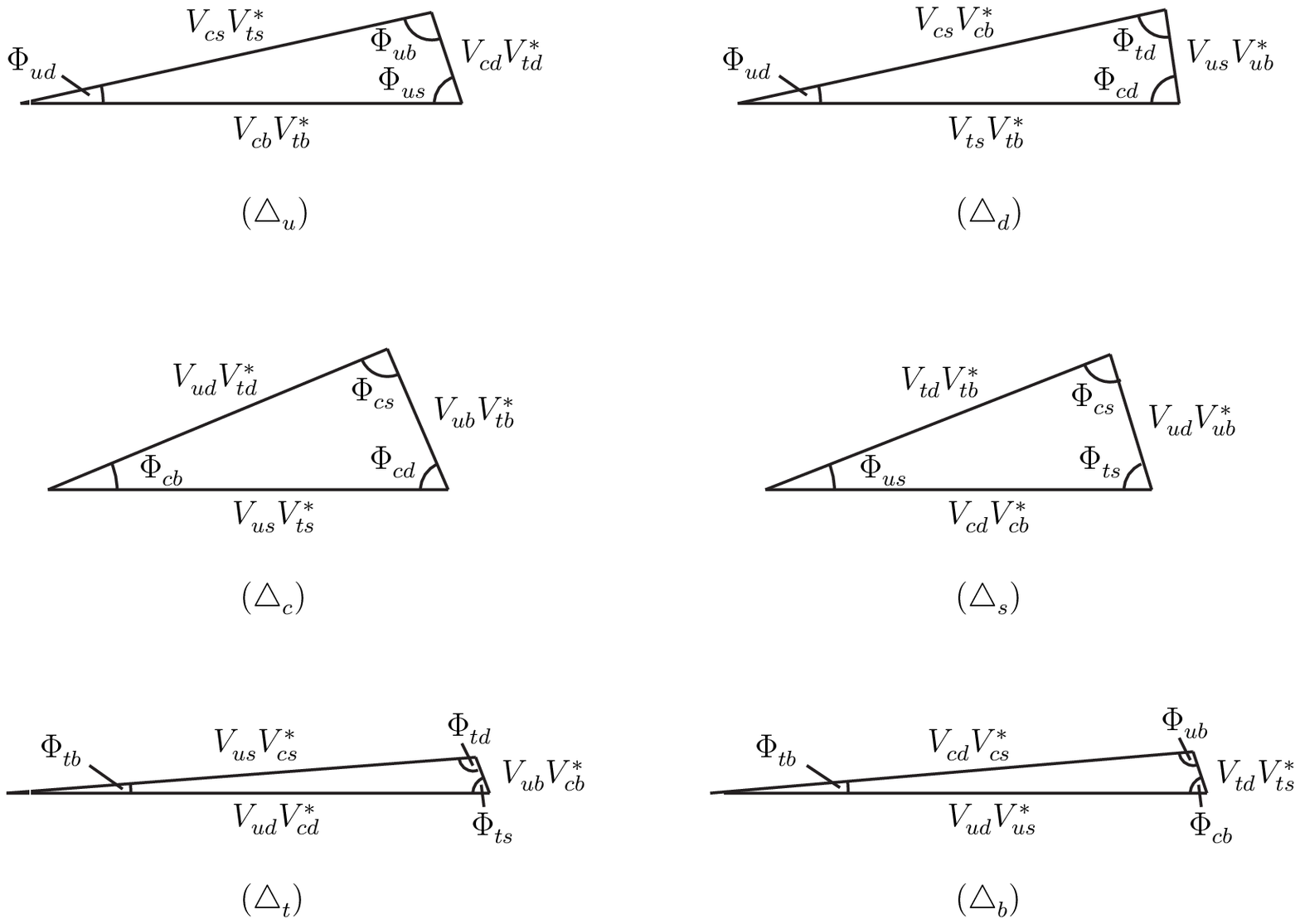}
\vspace{-7.5cm}\caption{Schematic diagrams for six UTs of the CKM
matrix in the complex plane, where each triangle is named by the
index that does not manifest in its three sides.} \vspace{1.4cm}
\end{center}
\end{figure}

\begin{figure}
\begin{center}
\vspace{6cm}
\includegraphics[bbllx=6.5cm, bblly=6.0cm, bburx=15.0cm, bbury=14.2cm,%
width=7cm, height=7cm, angle=0, clip=0]{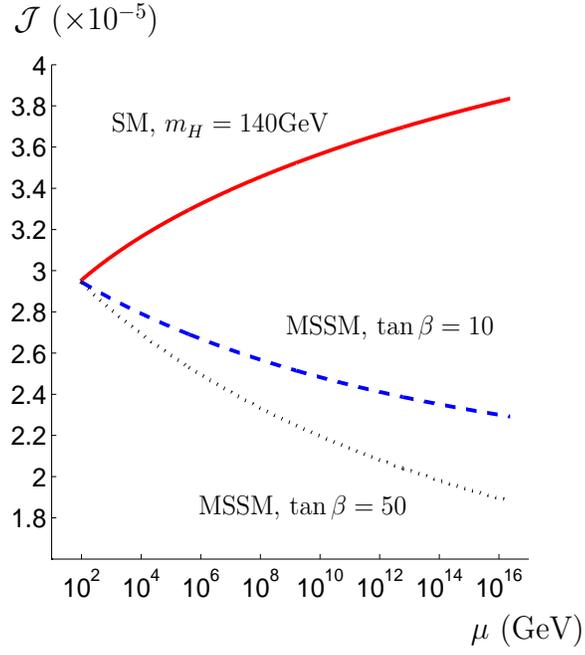}
\vspace{-3.2cm}\caption{Running behaviors of the Jarlskog invariant
$\cal J$ from $M^{}_Z$ to a superhigh-energy scale in the SM or in
the MSSM, where ${\cal J} = 2.95 \times 10^{-5}$ at $M_Z$ has been
input.} \vspace{1.0cm}
\end{center}
\end{figure}

\begin{table}[h]
\caption{Radiative corrections to the sides of six UTs at $\Lambda \sim 10^{14}$ GeV in the SM or MSSM.}
\begin{center}
\begin{tabular}{c|c|c|c|c|c}
\hline
\multicolumn{2}{c|}{} & & \multicolumn{3}{c}{$\Lambda \sim 10^{14}$ GeV} \\
\cline{4-6}
\multicolumn{2}{c|}{~~~~~~~~~~~~~~~~~~} & ~~~~~~ $M^{}_{Z} $ ~~~~~~ & SM ($m^{}_{H} = 140$ GeV) & MSSM ($\tan\beta = 10$) & MSSM ($\tan\beta = 50$) \\
\hline
& $|V^{}_{cd} V^{*}_{td}|$ & $1.95 \times 10^{-3}$ & $2.19 \times 10^{-3}$ & $1.74 \times 10^{-3}$ & $1.59 \times 10^{-3}$ \\
\cline{2-6}
$\triangle^{}_{u}$ & $|V^{}_{cs} V^{*}_{ts}|$ & $3.94 \times 10^{-2}$ & $4.44 \times 10^{-2}$ & $3.52 \times 10^{-2}$ & $3.23 \times 10^{-2}$ \\
\cline{2-6}
& $|V^{}_{cb} V^{*}_{tb}|$ & $4.12 \times 10^{-2}$ & $4.65 \times 10^{-2}$ & $3.68 \times 10^{-2}$ & $3.38 \times 10^{-2}$ \\
\hline
& $|V^{}_{ud} V^{*}_{td}|$ & $8.40 \times 10^{-3}$ & $9.47 \times 10^{-3}$ & $7.50 \times 10^{-3}$ & $6.89 \times 10^{-3}$ \\
\cline{2-6}
$\triangle^{}_{c}$ & $|V^{}_{us} V^{*}_{ts}|$ &  $9.14 \times 10^{-3}$ & $1.030 \times 10^{-2}$ & $8.16 \times 10^{-3}$ & $7.49 \times 10^{-3}$ \\
\cline{2-6}
& $|V^{}_{ub} V^{*}_{tb}|$ & $3.51 \times 10^{-3}$ & $3.96 \times 10^{-3}$ & $3.13 \times 10^{-3}$ & $2.88 \times 10^{-3}$ \\
\hline
& $|V^{}_{ud} V^{*}_{cd}|$ & 0.219 & 0.220 & 0.220 & 0.220 \\
\cline{2-6}
$\triangle^{}_{t}$ & $|V^{}_{us} V^{*}_{cs}|$ & 0.219 & 0.220 & 0.220 & 0.220 \\
\cline{2-6}
& $|V^{}_{ub} V^{*}_{cb}|$ & $1.45 \times 10^{-4}$ & $1.84 \times 10^{-4}$ & $1.16 \times 10^{-4}$ & $0.97 \times 10^{-4}$ \\
\hline
& $|V^{}_{us} V^{*}_{ub}|$ & $7.93 \times 10^{-4}$ & $8.94 \times 10^{-4}$ & $7.08 \times 10^{-4}$ & $6.50 \times 10^{-4}$ \\
\cline{2-6}
$\triangle^{}_{d}$ & $|V^{}_{cs} V^{*}_{cb}|$ & $4.02 \times 10^{-2}$ & $4.53 \times 10^{-2}$ & $3.59 \times 10^{-2}$ & $3.29 \times 10^{-2}$ \\
\cline{2-6}
& $|V^{}_{ts} V^{*}_{tb}|$ & $4.05 \times 10^{-2}$ & $4.56 \times 10^{-2}$ & $3.61 \times 10^{-2}$ & $3.32 \times 10^{-2}$ \\
\hline
& $|V^{}_{ud} V^{*}_{ub}|$ & $3.42 \times 10^{-3}$ & $3.86 \times 10^{-3}$ & $3.06 \times 10^{-3}$ & $2.80 \times 10^{-3}$ \\
\cline{2-6}
$\triangle^{}_{s}$ & $|V^{}_{cd} V^{*}_{cb}|$ &  $9.30 \times 10^{-3}$ & $1.05 \times 10^{-2}$ & $8.31 \times 10^{-3}$ & $7.63 \times 10^{-3}$ \\
\cline{2-6}
& $|V^{}_{td} V^{*}_{tb}|$ & $8.62 \times 10^{-3}$ & $9.71 \times 10^{-3}$ & $7.69 \times 10^{-3}$ & $7.06 \times 10^{-3}$ \\
\hline
& $|V^{}_{ud} V^{*}_{us}|$ & 0.220 & 0.220 & 0.220 & 0.220 \\
\cline{2-6}
$\triangle^{}_{b}$ & $|V^{}_{cd} V^{*}_{cs}|$ & 0.220 & 0.219 & 0.220 & 0.220 \\
\cline{2-6}
& $|V^{}_{td} V^{*}_{ts}|$ & $3.49 \times 10^{-4}$ & $4.44 \times 10^{-4}$ & $2.78\times 10^{-4}$ & $2.35 \times 10^{-4}$ \\
\hline
\end{tabular}
\end{center}
\end{table}

\end{document}